\documentclass[conference,review]{IEEEtran}
\IEEEoverridecommandlockouts
\usepackage{cite}
\usepackage{amsmath,amssymb,amsfonts}
\usepackage{algorithmic}
\usepackage{graphicx}
\usepackage{textcomp}
\usepackage{xcolor}
\usepackage{multirow}
\usepackage{array}
\usepackage{makecell}
\usepackage{tabularray}
\usepackage{lipsum}

\def\BibTeX{{\rm B\kern-.05em{\sc i\kern-.025em b}\kern-.08em
    T\kern-.1667em\lower.7ex\hbox{E}\kern-.125emX}}
\begin{document}

\title{ENOVA: Autoscaling towards Cost-effective and Stable Serverless LLM Serving
}

\author{
\IEEEauthorblockN{Tao Huang}
\IEEEauthorblockA{\textit{EmergingAI} \\
Hong Kong, China \\
shendu-ht@emergingai-tech.com} \\

\IEEEauthorblockN{Zachary Bright}
\IEEEauthorblockA{\textit{Exascale Labs} \\
USA \\
zack@exascalelabs.org}

\and

\IEEEauthorblockN{Pengfei Chen}
\IEEEauthorblockA{\textit{School of Computer Science and Engineering} \\
{Sun Yat-Sen University, China} \\
chenpf7@mail.sysu.edu.cn} \\

\IEEEauthorblockN{Wenxin Xie, Kecheng Huang}
\IEEEauthorblockA{\textit{EmergingAI} \\
Hong Kong, China \\
{wenxinxie,kecheng}@emergingai-tech.com}

\and

\IEEEauthorblockN{Kyoka Gong, Jocky Hawk}
\IEEEauthorblockA{\textit{EmergingAI} \\
Hong Kong, China \\
{kyokagong,jockyhawk}@emergingai-tech.com} \\

\IEEEauthorblockN{Zhi Ji}
\IEEEauthorblockA{\textit{Division of Engineering Science} \\
University of Toronto, Canada \\
whitney.ji@mail.utoronto.ca}
}

\maketitle

\begin{abstract}
Since the increasing popularity of  large language model (LLM) backend  systems, it is common and necessary to deploy stable serverless serving of LLM on multi-GPU clusters with autoscaling. However, there exist challenges because the diversity and co-location of applications in multi-GPU clusters will lead to low service quality and GPU utilization. To address them, we build ENOVA, a deployment, monitoring and autoscaling service towards serverless LLM serving. ENOVA deconstructs the execution process of LLM service comprehensively, based on which ENOVA designs a configuration recommendation module for automatic deployment on any GPU clusters and a performance detection module for autoscaling. On top of them, ENOVA implements a deployment execution engine for multi-GPU cluster scheduling. The experiment results show that ENOVA significantly outperforms other state-of-the-art methods and is suitable for wide deployment in large online systems. 
\end{abstract}


\section{Introduction}

The rising emergence of large language models (LLMs) like GPT~\cite{achiam2023gpt} has spawned many new applications in various domains such as programming~\cite{luo2023wizardcoder,li2023starcoder}, academic~\cite{blecher2023nougat}, legal~\cite{cui2023chatlaw}, medical~\cite{thirunavukarasu2023large} assistants and embodied agents~\cite{wang2023describe}, which improve the efficiency of our work and routines. These applications  generally run on GPU cloud in industrial environments, where an LLM request can be $10\times$ more expensive than a traditional query~\cite{kwon2023efficient}. Hence, previous researchers focus on optimizing LLM inference so that to increase the throughput and reduce the cost of LLM serving.

Many approaches~\cite{dao2022flashattention, yu2022orca, kwon2023efficient,rasley2020deepspeed,shoeybi2019megatron,dettmers2022llm,frantar2022gptq,xiao2023smoothquant} have been proposed to optimize LLM inference. These techniques can be classified into five general categories including computation~\cite{shoeybi2019megatron,rasley2020deepspeed}, weight or activation quantization~\cite{dettmers2022llm,frantar2022gptq,xiao2023smoothquant}, memory management~\cite{kwon2023efficient}, kernel operator optimization~\cite{dao2022flashattention,rasley2020deepspeed}, and batching requests~\cite{yu2022orca}. However, few of them focus on stable and auto-scaled serverless LLM serving in  multi-GPU clusters. There are still challenges to provide a scheduling service so that LLM developers no longer need to spend time providing stable and scalable LLM services. The challenges are shown as follows.
\begin{itemize}
    \item \textbf{Unaware of the diversity of LLM tasks.} Taking the LLM agent as an example, the agent can contain a profiling module to identify its role, a memory module to place it into a dynamic environment, and a planning module to determine future actions. There is a significant difference in the context length of LLM tasks in this agent. This diversity of task targets in different modules will affect the performance of LLM service by the input and output sequence length. 
    \item \textbf{Less focus on deployment in distributed multi-GPU clusters.} The GPU clusters are mainly from multiple regions with various types of GPU devices in industrial environments. There are differences in computing capabilities among various GPU devices. If without a guidance on multi-cluster deployment of LLM service, the performance of LLM service will be impacted due to this computing capacity difference.
    \item \textbf{Unable to auto-scale LLM service.} When an LLM service is deployed online, the requests issued to the LLM service are complex and non-stationary. Without autoscaling, the service quality degrades when the number of requests surges, while the resource waste exists when it drops sharply.
\end{itemize}

To address these challenges, we introduce \textbf{ENOVA}, a service designed to enhance the stability and cost-efficiency of serverless LLM serving through deployment, monitoring, and autoscaling capabilities. ENOVA comprises two primary modules: the service configuration module, which determines the optimal configurations for LLM services, and the performance detection module, tasked with monitoring service quality and resource utilization and identifying anomalies that necessitate autoscaling actions. On top of them, ENOVA integrates a comprehensive execution engine that manages deployment, monitoring, and autoscaling across multi-GPU clusters. Our contributions are summarized as follows:

\begin{itemize}
    \item We have designed a service configuration recommendation module that is designed for LLMs deployed across any GPU cluster. This module is also adaptable to a variety of application agents, ensuring optimal configurations and service quality in diverse environments. Notably, this approach is scalable, supporting multi-agent systems and multi-GPU deployments effectively.
    \item We have developed a performance detection module that continuously monitors service quality and resource utilization in real-time. This module is essential to ensuring the automatic adjustment of online-deployed LLM services, thus maintaining optimal service reliability. 
    \item We have implemented a comprehensive service for the deployment, monitoring, and autoscaling of LLMs. This service significantly reduces the workload of LLM developers by providing stable and scalable performance across multi-GPU clusters. The implementation code is publicly available for further research and development at https://github.com/Emerging-AI/ENOVA.
    \item We evaluate ENOVA for various scenarios and compare it with several state-of-the-art systems. The experiment results demonstrate that ENOVA outperforms the baseline approaches. 
    
\end{itemize}


\section{Background}\label{sec:2}

\subsection{GPU Scheduling}

A graphics processing unit (GPU) is a specialized device that consists of several \textit{streaming multiprocessors} (SMs). 
Multiple user contexts can  run on one SM concurrently, each of which is allocated a static share of local resources. A typical user context will consist of numerous \textit{kernels}, which are executed sequentially. The applications to process user contexts typically interact with the GPU at the kernel level. A kernel function  is  scheduled to run parallelly in blocks of \textit{threads} within a grid, where thread is the smallest unit of GPU computation in the CUDA framework.

The GPU scheduling is to submit kernels to hardware queues within GPUs and schedule them to target SMs with available resources. From the macro perspective, the LLM inference tasks can be refined into tens types of kernels, such as \textit{"aten::add"} and \textit{"aten::transpose"} in \textit{PyTorch}. Each type of kernel is corresponding to one basic operations defined in PyTorch, for instance the kernel \textit{"aten::add"} is responsible for executing function \textit{"torch.add"}. Eventually, the LLM inference tasks are scheduled and executed at the kernel level. Model serving systems like NVIDIA's Triton~\cite{triton} and Ray Serve~\cite{ray-serve} are responsible for receiving requests and dispatching LLM inference executions to target GPUs.

\subsection{LLM Inference}

LLMs are always autoregressive transformer-based generative models~\cite{vaswani2017attention}, which predict the next token based on past tokens. Since LLMs generate one token at a time, the LLM inference  generates subsequent tokens iteratively until they trigger stopping criteria. Within one iteration, the input tokens are transmitted into the model which contains several transformer layers with one attention, two layer norm, and two feed-forward layers. Essentially, the attention layer uses past tokens to generate the next token. The output $o_i$ of the attention layer at the $i^{th}$ iteration can be express as: 
\begin{align}
    o_i & = \sum_{j=1}^{i} a_{i,j} v_{j}  \\ 
    a_{i,j} & = \frac{\exp(q_i^T k_j / \sqrt{d_h})}{\sum_{t=1}^{i} \exp(q_i^T k_t / \sqrt{d_h})} 
\end{align}
where $k_t$ is the key at $t^{th}$ iteration, $v_j$ is the value at $j^{th}$ iteration, $a_{i,j}$ is the attention weight of $v_j$ at $i^{th}$ iteration, $q_i$ is the query at $i^{th}$ iteration, and $d_h$ is the hidden dimension of model. The output generation at each iteration relies on the previous generated key and value vectors.

To avoid the redundant computation in attention layer, the key and value vectors of past tokens are often cached for generating the next tokens, which are known as \textit{KV cache}~\cite{pope2023efficiently}. Thus, the LLM inference involves two phases, namely the prefill phase and autoregressive generation phase. In the prefill phase, the LLM takes the user context tokens as input and generates the first token, during which the intermediate keys and values are prefilled in KV cache. In the autoregressive generation phase, the LLM takes the previous generated token as input and generates the following  tokens at a time, until one stopping criteria is triggered. 

This inference mode results in at least two perspectives that need to be optimized. The first is the GPU memory requirements caused by KV cache, which is linearly growing with batch size and sequence length. Eventually, it leads to the limitation of the throughput and long-context inputs of LLM Service. The second is the low GPU utilization brought by sequential autoregressive generation, which will lead to lower throughput of LLM Service.

\subsection{Optimizing the LLM Inference}

To optimize the LLM inference, several approaches~\cite{dao2022flashattention, yu2022orca, kwon2023efficient,rasley2020deepspeed,shoeybi2019megatron,dettmers2022llm,frantar2022gptq,xiao2023smoothquant} has been proposed. For instance,
PagedAttention~\cite{kwon2023efficient} is an attention algorithm which optimizes the KV cache management by storing attention keys and values in non-contiguous paged memory. It achieves near-zero waste in KV cache memory and flexible sharing of KV cache within and across requests.
FlashAttention~\cite{dao2022flashattention} is an IO-aware exact attention algorithm which optimizes the LLM inference computation by reducing the number of memory reads/writes between GPU high bandwidth memory (HBM) and GPU on-chip SRAM. 
OCRA~\cite{yu2022orca} is an iteration-level scheduling mechanism which addresses the multi-iteration issues in autoregressive generation phase. It adopts selective batching to batch arbitrary requests processing tokens at different positions so that the scheduler can execute at the granularity of iteration instead of request.  
SpecInfer~\cite{miao2023specinfer} is an LLM serving system which focuses on accelerating generative LLM inference with speculative inference and token tree verification. It combines various boost-tuned small language models to jointly predict the LLM outputs and uses an LLM as a token tree verifier, so that to reduce the end-to-end latency and computational requirement. 
Despite many researchers are focusing on optimizing the LLM inference, there are still several practical challenges to be addressed from the perspective of serverless LLM serving, which will be demonstrated in the next section.




\section{Motivation}\label{sec:3}

\subsection{Problem Formulation}\label{sec:3.1}

Instead of model inference, this paper aims at serverless LLM serving so that LLM developers can focus on model training or fine-tuning and no longer need to spend time providing stable and scalable LLM service in multi-GPU clusters. To achieve that, there are two target problems to be addressed.

\begin{itemize}
    \item Given a pre-trained or fine-tuned LLM, we need to determine the optimal configurations so that to maximize interactive performance and minimize costs. Then the LLM can be dispatched to the target idle machines in multi-GPU clusters. 
    \item Given an online LLM service, we need to observe the service quality in real time. When we detect the anomalous resource utilization or service quality, we need to redetermine the optimal configuration of LLM service and re-schedule it automatically.
\end{itemize}

\subsection{Goals of Serverless LLM Serving}\label{sec:3.2}

To effectively address the two target problems and achieve serverless LLM serving, several specific objectives need to be accomplished. Firstly, it's important to recognize that LLM services require distinct service configurations when extended to different application agents. For example, the configuration termed as maximal number of sequences (also \textit{max\_num\_seqs}) is crucial, as it determines the number of requests that can be handled simultaneously by the LLM service. This requires allocating sufficient GPU memory to handle multiple requests in parallel. Diverse agents may have varied short-term and long-term memory structures that are critical for executing future actions. These differences directly impact the input sequence length required for LLM tasks. During the LLM generation phase, the greater the input sequence length, the more GPU memory is necessary. However, it is important to point out that there is a maximum memory capacity for each GPU, which will limit the \textit{max\_num\_seqs} by the context length of LLM tasks. In practice, many service configurations will be affected when an LLM service is extended to different application agents. We need to recommend the optimal service configurations for each application agent, which can be considered as the \textbf{multi-agent deployment goal} of serverless LLM serving.

Secondly, it's crucial to determine the optimal service configurations for deploying an LLM on multi-GPU clusters. Still using the \textit{max\_num\_seqs} as an example, the capacity to process multiple requests concurrently is directly proportional to the computing capacity of the GPUs deployed. For instance, the NVIDIA H100 80 GB can process more requests concurrently compared to the GeForce RTX 4090 24 GB. Therefore, the \textit{max\_num\_seqs} for an LLM service deployed on the NVIDIA H100 80 GB should be set higher to fully utilize its superior computing capabilities. Due to differences in computing capacities across various GPU devices, the service configurations of an LLM service will be significantly affected. We need to recommend the optimal service configurations for the LLM deployed on multi-GPU clusters, which can be understood as the \textbf{multi-GPU deployment goal} of serverless LLM serving.

\begin{figure}
    \centering
    \includegraphics[width=\linewidth]{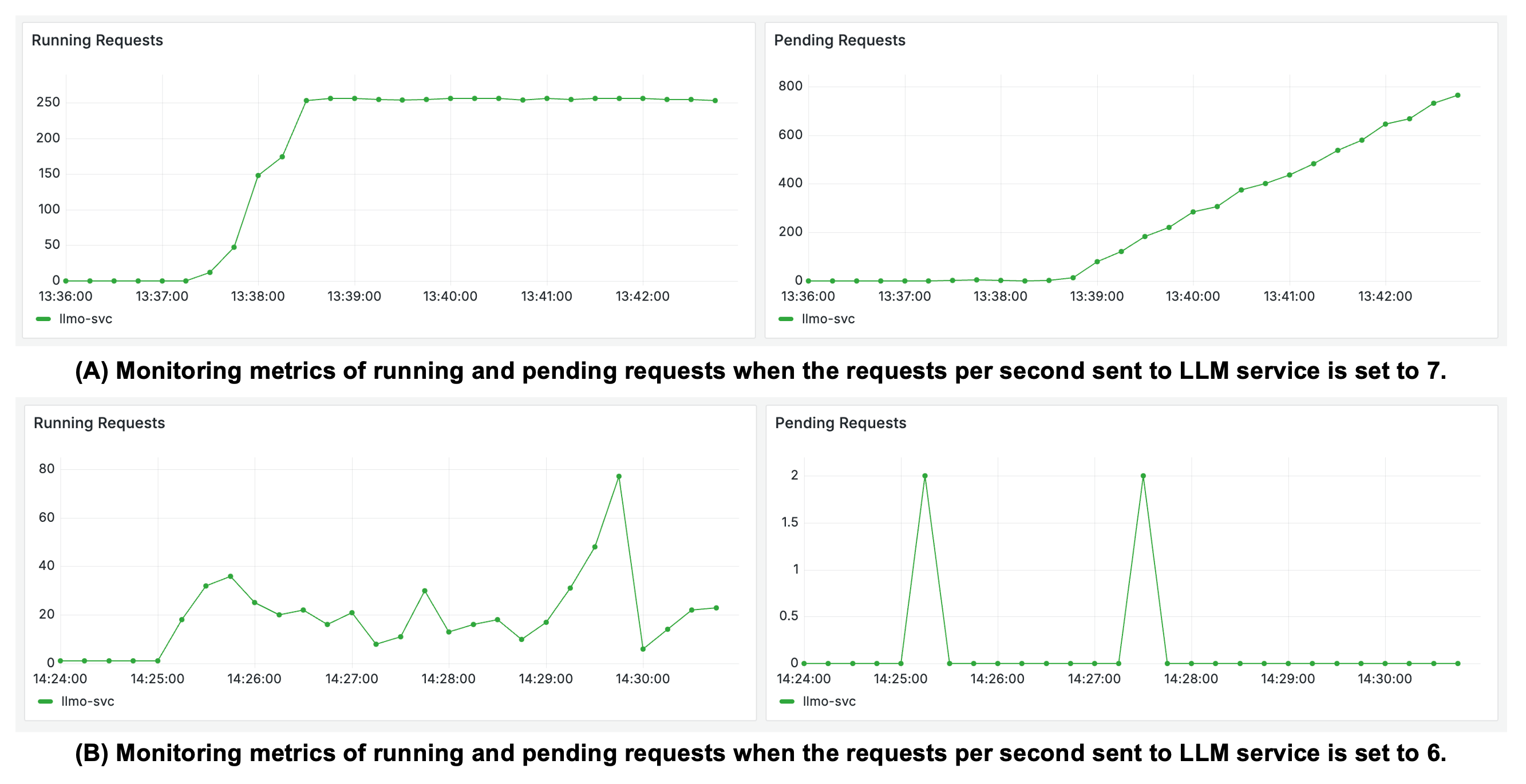}
    \caption{The monitoring metrics for running and pending requests when the requests per second sent to LLM service is set to $7$ and $6$ respectively.}
    \label{fig:auto-scaling-goal}
\end{figure}

Thirdly, it is essential to point out that there is an upper limit to the number of requests per second that can be sent to an LLM service. The previous two goals focus on increasing this limit using fewer GPU resources when LLM services are deployed on multi-GPU clusters and extended to various application agents. When the requests per second slightly exceed this limit, the LLM service will be down, causing all incoming requests significant delays until the HTTP connection times out. Fig.~\ref{fig:auto-scaling-goal} is a practical case in an industrial environment. Fig.~\ref{fig:auto-scaling-goal} (A) and (B) respectively presents the monitoring metrics for running and pending requests when the requests per second sent to LLM service are set to $7$ and $6$. The running requests represent the number of requests currently running on the GPU, while the pending requests are those queued and waiting to be processed. We can observe that when requests per second are set to $6$, all requests can be processed without any delays caused by waiting in queue. However, when requests per second increase to $7$, more and more requests accumulate and pend in the queue after the number of running requests reaches \textit{max\_num\_seqs}. To handle these pending requests, we need to add another replica of the LLM service. In industrial environments, the requests per second are not constant but hard to predict, varying with demands and user activity. To provide a stable and scalable service, it's essential to implement automatic scaling of the LLM service. This can be regarded as the \textbf{autoscaling goal} of serverless LLM serving.

\begin{figure*}
    \centering
    \includegraphics[width=\linewidth]{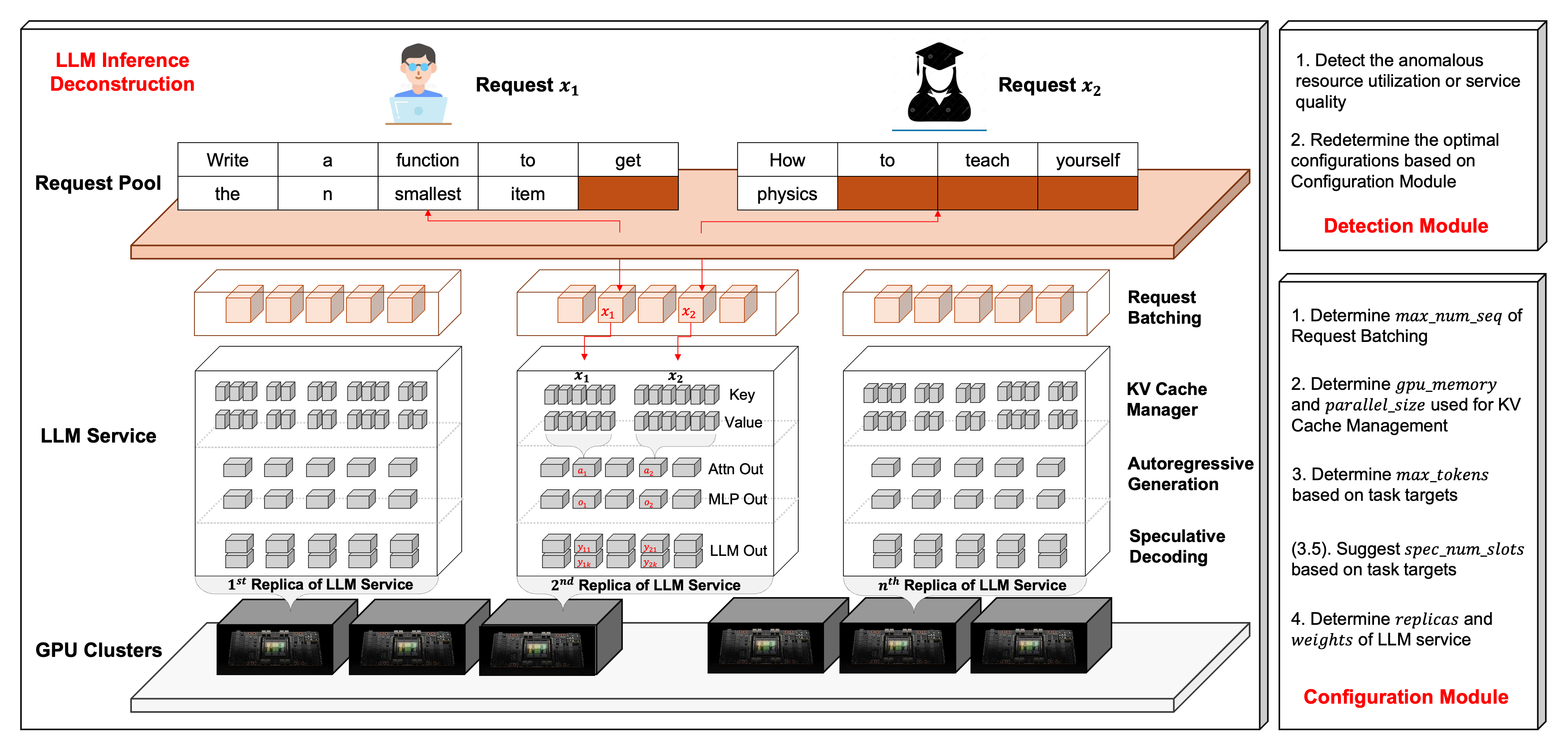}
    \caption{The procedure by which we deconstruct the process of LLM inference led to the design of configuration module and detection module in ENOVA. }
    \label{fig:ENOVA}
\end{figure*}

\section{Approach Overview}\label{sec:4}

To achieve the above goals, To address these challenges, we propose ENOVA, a deployment, monitoring and autoscaling service towards stable and cost-effective serverless LLM serving. Fig.~\ref{fig:ENOVA} illustrates how ENOVA deconstructs the execution process of LLM inference. The request pool collects user inputs and dispatches them to a replica of the LLM service deployed on GPU clusters via an HTTP load balancer. For continuously incoming requests, the LLM service merges these newly coming requests with previously running requests into a batch via continuous batching~\cite{yu2022orca}. During the autoregressive generation phase, the key and value vectors of all running requests are stored as non-contiguous blocks using PagedAttention~\cite{kwon2023efficient} and leveraged to generate attention output. When generating the output in a single step, speculative decoding strategies such as SpecInfer~\cite{miao2023specinfer} are utilized to predict the next $k$ tokens, thereby accelerating the LLM inference process. These inference techniques introduce a set of service configurations that must be carefully configured to provide a stable and cost-effective LLM service. Therefore, ENOVA designs two key modules: the service configuration module and the performance detection module. The configuration module aims to determine the optimal configurations for LLM services, including the \textit{max\_num\_seqs} for continuous batching and \textit{gpu\_memory} for PagedAttention. Meanwhile, the detection module is responsible for monitoring service quality and resource utilization, identifying anomalies that trigger autoscaling actions. In this section, we will provide a detailed exploration of these modules and their respective functionalities.

\subsection{Service Configuration Module}

To achieve the \textbf{multi-agent deployment} and \textbf{multi-GPU deployment goals}, ENOVA has designed a service configuration recommendation module. This module is built for LLMs deployed on any GPU cluster and extended to various application agents, ensuring optimal configurations and performance across diverse environments. Taking \textit{gpu\_memory} allocated for LLM service as an example, the necessary memory allocation can be expressed as:
\begin{align}
    mem \approx & {\ } params * sizeof(dtype) {\ } + max\_num\_seqs\ * \nonumber \\ 
     & seq\_length * token\_mem + others,  \label{eq:mem}
\end{align}
where $param$ and $token\_mem$ is the parameter size and memory of KV cache defined by deployed LLM, $seq\_length$ is the context length of user input depended on agent tasks, and $max\_num\_seqs$ is specified by the computing capacity of GPU devices. Due to the diversity of LLM agent tasks and differences in GPU devices, the memory requirements can vary significantly, necessitating a flexible configuration approach to optimize resource allocation and ensure efficient performance across all deployments. Therefore, it is essential to design the methodology from a scalable perspective, rather than relying on heuristic calculations such as those described in equation~\ref{eq:mem}. In this paper, ENOVA designs and implements a monitoring system which will be illustrated in section~\ref{sec:5}, and determines the service configurations based on the monitoring metrics. Since the diversity of LLM agent tasks and GPU devices can be reflected in the variations in metric observations, this approach is suitable for multi-agent and multi-GPU deployment.

Fig.~\ref{fig:ENOVA} depicts the process employed by the service configuration module in ENOVA to establish configurations. These configurations are meticulously derived from the deconstruction of LLM inference processes, which are critical for optimizing resource utilization and ensuring service stability. Detailed information regarding the source layer, description, and objectives of these configurations is systematically presented in TABLE~\ref{tab:config}. In the remaining of this section, we will elaborate on how ENOVA determines these service configurations.

\begin{table*}
    \centering
    \caption{The involved service configurations of ENOVA.}
    \label{tab:config}
    \renewcommand\arraystretch{1.2}
    \begin{tabular} {cccc}
    \hline
    \textbf{Source Layer} & \textbf{Configuration} & \textbf{Description} 
            & \textbf{Objective} \\
    \hline
    LLM Deployment & \makecell[c]{\textit{parallel\_size} \\ \textit{gpu\_memory}} 
            & \makecell[c]{{the tensor or pipeline parallel size when deploying LLM} \\ {the allocated GPU memory for deploying LLM}} 
            & {Lower Cost, Service quality} \\
    LLM Inference & \textit{max\_num\_seqs} 
            & the maximal number of sequences handled simultaneously 
            & {Lower Cost, Service quality} \\
    LLM API Service & \textit{max\_tokens} 
            & the maximal number of output tokens of user request
            & Service quality \\
    LLM Load Balancer & \makecell[c]{\textit{replicas} \\ \textit{weights}} 
            & \makecell[c]{{the number of deployed LLM services} \\ {the routing weights that new requests dispatched to the target replica}} 
            & Lower Cost, Service quality \\
    \hline
    \end{tabular}
\end{table*}

\subsubsection{Determine \textit{max\_num\_seqs}}

This fundamental configuration specifies the maximal number of sequences that can be handled concurrently, which can directly influences the allocation of GPU memory resources and execution time per batch output token. We determine \textit{max\_num\_seqs} based on the following equation: 
\begin{equation}
    max\_num\_seqs \approx n_{limit} \times t^r_{limit}
\end{equation}
where $n_{limit}$ represents the maximal number of requests per second sent to an LLM service and $t^r$ denotes the corresponding execution time of user requests. This formula is straightforward, as it posits that the maximum number of sequences is determined by multiplying the number of requests per second by the average execution time of each request. Meanwhile, a fundamental assumption in this equation is that the maximum number of requests per second remains constant, irrespective of any adjustments made to the service configurations. This assumption will be verified and discussed in section~\ref{sec:7.1}. 

Then, we determine $n_{limit}$ and $t^r_{limit}$ based on the monitoring metrics of LLM service. TABLE~\ref{tab:max_num} presents the involved metrics, where $n^f$ represents the number of finished requests per unit time of an LLM service, $n^r$ specifies the number of running requests simultaneously per unit time, and $t^r$ denotes the execution time per user request. Theoretically, $n_{limit}$ is the upper limit of $[n_{t-w}^{f}, \cdots n_{t}^{f}]$, and $t^r_{limit}$ is the execution time $t^r$ corresponding to when $n^f$ reaches $n_{limit}$. However, in an industrial environment, the observed values of $n^f$ can deviate significantly from $n_{limit}$. Therefore, we first identify whether $n^f$ deviate from $n_{limit}$ by modeling the following function.
\begin{equation}
    n^f = f(n^r)
\end{equation}
If there is no significant relationship between $n^f$ and $n^r$, it can be inferred that $n^f$ has reached $n_{limit}$; if not, the opposite is deduced. In this paper, ordinary least squares regression (OLS)~\cite{ramsey1969tests} is used to model $n^f$ and $n^r$, and t-test is adopted to determine whether there is a significant relationship between $n^f$ and $n^r$. If $n^f$ deviates from $n_{limit}$, we can deduce that $n_{limit}$ and $t^r_{limit}$ are generated from extreme value distributions of $[n_{t-w}^{f}, \cdots n_{t}^{f}]$ and $[t_{t-w}^{r}, \cdots t_{t}^{r}]$. If not, $n_{limit}$ and $t^r_{limit}$ are generated from normal distribution. Eventually, we estimate the extreme value distribution or normal distribution with kernel density estimate (KDE)~\cite{parzen1962estimation}, based on which $n_{limit}$ and $t^r_{limit}$ are determined.

\begin{table}
    \centering
    \caption{The part of monitoring metrics in ENOVA.}
    \label{tab:max_num}
    \renewcommand\arraystretch{1.2}
    \begin{tabular} {ccc}
    \hline
    \textbf{Metric Description} & \textbf{Symbol} & \textbf{Observations} \\
    \hline
    Number of finished requests per unit time & $n^f$ & $[n_{t-w}^{f}, \cdots n_{t}^{f}]$ \\
    Number of running requests per unit time & $n^r$ & $[n_{t-w}^{r}, \cdots n_{t}^{r}]$ \\
    Number of arriving requests per unit time & $n^a$ & $[n_{t-w}^{a}, \cdots n_{t}^{a}]$ \\
    Number of pending requests per unit time & $n^p$ & $[n_{t-w}^{p}, \cdots n_{t}^{p}]$ \\
    Execution time per user request & $t^r$  & $[t_{t-w}^{r}, \cdots t_{t}^{r}]$ \\
    GPU memory utilization & $m^{u}$ & $[m_{t-w}^{u}, \cdots m_{t}^{u}]$ \\
    GPU utilization & $g^{u}$ & $[g_{t-w}^{u}, \cdots g_{t}^{u}]$ \\
    \hline
    \end{tabular}
\end{table}

\subsubsection{Determine \textit{parallel\_size} and \textit{gpu\_memory}}

These configurations define the allocated GPU memory for LLM service. If the GPU memory allocated to an LLM service is insufficient, the service quality will be negatively affected due to inadequate memory for storing the KV cache of requests. Conversely, excessive allocation of GPU memory results in resource waste. Therefore, it is essential to meticulously determine this configuration. In the process of determining this configuration, we meticulously analyze the relationship between the GPU memory utilization, denoted as $m^{u}$, and number of running requests per unit time, $n^{r}$, as presented in TABLE~\ref{tab:max_num}. The relationship is formulated as follows.
\begin{align}
    m^{u} & = g(n^r) \nonumber \\
    gpu\_memory & \approx g(max\_num\_seqs)
\end{align}
We then determine the required $gpu\_memory$ by setting $n^r$ equal to $max\_num\_seqs$. This formula is straightforward to derive, as the variance in $gpu\_memory$ can be attributed to changes in $n^r$. Specifically, an increase in the number of running requests per unit time corresponds to greater utilization of GPU memory used for KV cache storage. Therefore, we can calculate the maximal allocated GPU memory for an LLM service by setting the number of running requests $n^r$ equal to $max\_num\_seqs$. In this paper, ordinary least squares
regression (OLS)~\cite{ramsey1969tests} is also adopted to model $m^{u}$ and $n^{r}$, based on which $gpu\_memory$ is determined and $parallel\_size$ can be derived based on existing hardware specifications.

\subsubsection{Determine \textit{max\_tokens}}

This configuration specifies the maximal number of output tokens of each user request. We manage this configuration because not all prompts are optimally designed. Poorly designed prompts may result in output tokens reaching their maximum limit, thereby degrading the quality of service. Appropriately limiting the maximum number of output tokens is crucial for maintaining the service quality. To determine the $max\_tokens$, we operate under the assumption that the number of output tokens for each type of LLM task is predetermined and should not deviate significantly from the established range. We will further discuss it in section~\ref{sec:7.2}. 

Then, we determine the $max\_tokens$ of each type of LLM task based on task clustering. ENOVA first adopts bge-large-en~\cite{bge_embedding} to convert user request text into embedding representations. Second, ENOVA adopts community detection algorithm ~\cite{fortunato2010community} to cluster the user requests by maximizing the modularity. ENOVA initially constructs a request graph by calculating the cosine similarity between embedding representations. Subsequently, ENOVA optimizes the objective function to identify communities within user requests, which can be described as follows:
\begin{equation}
    \mathcal{L}_{Q} = \frac{1}{2m} \sum_{i,j} \begin{bmatrix} A_{i,j} - \frac{k_i k_j}{2m} \end{bmatrix} \cdot \delta(i, j),
 \end{equation}
where $i$ and $j$ denote the $i^{th}$ and $j^{th}$ requests; $m$ is the total number of edges in the request graph; $k_i$ and $k_j$ represent the degrees of request $i$ and $j$; and $\delta(i, j)$ equals $1$ if $i$ and $j$ are in the same community and $0$ otherwise. Ultimately, user requests will be segmented into distinct communities based on their task targets. For each new request, ENOVA determines its community by calculating the cosine similarity between the request and the centroid of each existing community. For each community, ENOVA uses KDE~\cite{parzen1962estimation} to model the density function of the length of generated output tokens, based on which the $max\_tokens$ of each community are determined.

\subsubsection{Determine \textit{replicas} and \textit{weights}}

These configurations control the number of replicas of deployed LLM services and the routing weights used to route new requests to the appropriate replica. In industrial environments, GPU clusters are often distributed across multiple regions, featuring a variety of GPU device types. These devices exhibit significant disparities in computing capabilities, unlike traditional computing devices. When deploying two replicas of LLM services on two distinct GPU devices, the absence of a strategic routing plan for user requests can lead to inefficiencies. The more powerful GPU may underutilize its computing capacities, while the less capable GPU could result in diminished service quality. Therefore, it is essential to accurately determine the appropriate number of $replicas$ and their corresponding $weights$ for effective multi-GPU deployment.

Then we determine the $replicas$ of each type of GPU based on the following equations:
\begin{gather}
    \mathrm{min} \sum_{i} score^{i} * replicas^{i}  \\
    \left\{\begin{array}{cc} 
        \sum_{i} n_{limit}^{i} * replicas^{i} & \leq n^f \\
        parallal\_size^{i} * replicas^{i} & \leq N^i  \end{array} \right. \nonumber
\end{gather}
where $score^i$ represents the matching score of LLM services deployed on $GPU^i$; $replicas^{i}$ denotes the number of replicas of LLM services deployed on $GPU^i$; $n_{limit}^{i}$ specifies the maximal number of requests per second that can be processed by LLM services on $GPU^i$; $parallal\_size^{i}$ indicates the designated parallel size for LLM service deployed on $GPU^i$; and $N_i$ is the total number of GPU devices of $GPU^i$. The matching score is calculated based on the difference between the required $gpu\_memory$ and the total memory of $GPU^i$. Eventually, the number of replicas, $replicas^{i}$, is determined using linear programming~\cite{dantzig2002linear}. The \textit{weight} of replicas is directly specified by the estimated $n_{limit}^{i}$. The new request will be routed to the target GPU device based on these routing weights.

\subsection{Performance Detection Module}\label{sec:4.2}

To achieve the \textbf{autoscaling goal}, ENOVA has developed a performance detection module. This module monitors service quality and resource utilization in real-time, particularly important due to the complexity of user requests in industrial environments. Fig.~\ref{fig:ENOVA} illustrates the detailed procedure of performance detection module. ENOVA first identifies anomalies in resource utilization or service quality using the monitoring metrics presented in TABLE~\ref{tab:max_num}, then redetermine the optimal configurations using the previously established service configuration module. 

When developing the performance detection model, it is essential to consider the rarity of anomalous samples in industrial environments, which poses challenges for collecting sufficient data for supervised learning. Additionally, the performance of unsupervised models can deteriorate when contaminated with a few anomalous samples~\cite{huang2022semi}. To address these challenges, ENOVA utilizes a semi-supervised learning approach~\cite{van2020survey}, specifically adopting a Variational Auto-Encoder (VAE)~\cite{kingma2013auto} as the base model. This model hypothesizes that the observed normal performance metrics $m$ are generated from an unobserved multivariate Gaussian distribution $z$, while the anomalous performance data deviate from this distribution. The input metrics are normalized prior to being fed into the VAE. Then ENOVA optimizes the evidence of variational lower bound (ELBO) which can be expressed as follows.
\begin{align}
    \mathcal{L}_{vae} = 
    & \frac{1}{|D^M|} \sum_{(m_i, l_i) \in D^M} l_i \cdot \mathbb{E}_{q_{\phi}(z|m)} \log (p_{\theta}(m|z)) \nonumber \\
    & - (1 + l_i)/{2} \cdot \beta(k) \cdot KL(q_{\phi}(z|m_i) \| p_{\theta}(z)),
\end{align}
In this equation, $l_i \in (1, -1)$ represents the label of performance metrics, where $l_i = 1$ indicates that $m_i$ is normal. The coefficient $\beta(k)$ from PI control~\cite{higgins2016beta} is used during the $k^{th}$ training iteration to ensure convergence of the objective function. The term $\mathbb{E}_{q_{\phi}(z|m)} \log (p_{\theta}(m|z))$ represents the expectation component of ELBO, while $KL(q_{\phi}(z|m_i) \| p_{\theta}(z))$ denotes the KL-divergence component of ELBO. ENOVA uses these labels to manage the components, allowing the few anomalous samples to effectively define the distribution boundaries.

\begin{figure*}
    \centering
    \includegraphics[width=0.9\linewidth]{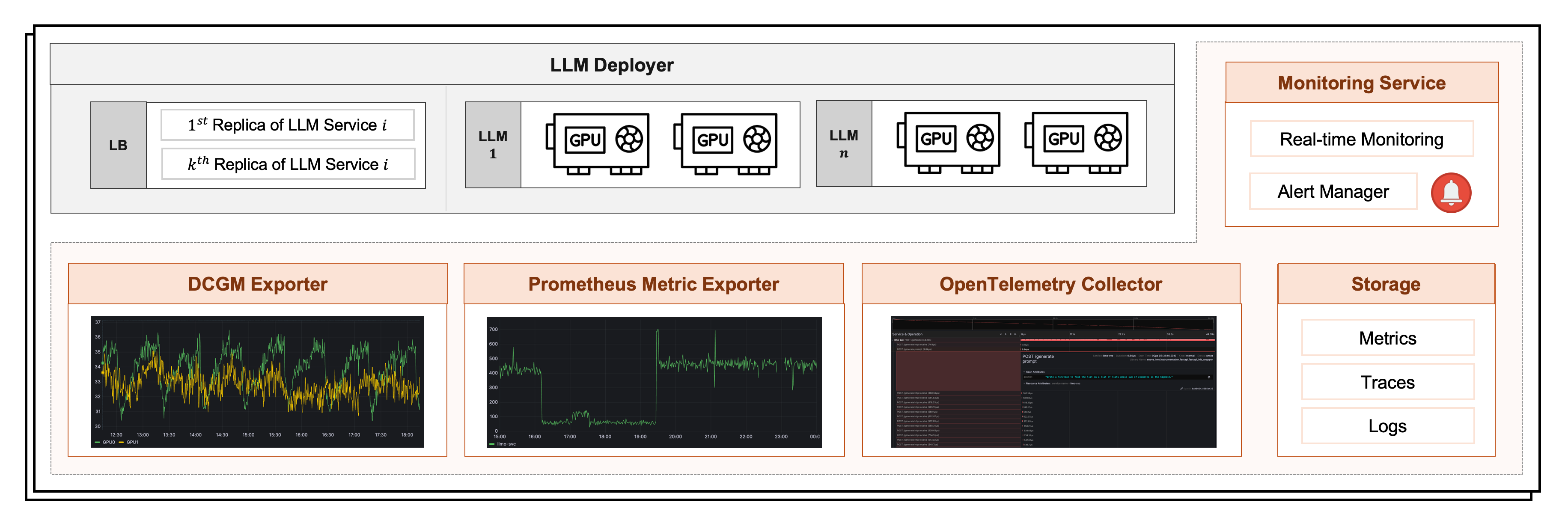}
    \caption{The implementation components of ENOVA, designed to ensure accurate execution of the deployment, monitoring, and autoscaling services.}
    \label{fig:Implement}
\end{figure*}

When utilizing the detection model to identify anomalous performance, ENOVA focuses on the KL-divergence where posterior distribution $q_{\phi}(z|m_i)$ is deviated from prior distribution $p_{\theta}(z)$. The threshold for detecting anomalies is automatically set using the peaks-over-threshold method~\cite{siffer2017anomaly}. An anomaly is detected if the KL-divergence for $m_i$ exceeds this threshold. Additionally, as the normalized values of input metrics can indicate either overload or underload, ENOVA uses the Mean Difference (MD) between the input metrics $m$ and the reconstructed metrics $m'$ to decide whether to scale up or down. Based on these, ENOVA then adjusts the service configurations using the service configuration module.

\section{System Implementation}\label{sec:5}

To ensure precise execution of the aforementioned methodologies, ENOVA has designed and implemented an advanced execution engine. This engine effectively manages deployment, monitoring, and autoscaling operations within multi-GPU clusters. Fig.~\ref{fig:Implement} illustrates the architecture of this engine, which is primarily composed of two main components: the LLM deployer and the monitoring system. In the subsequent sections, we provide a detailed description of each component, explaining their functionalities and interactions within the system.

The LLM deployer enables the automatic deployment of LLM services, utilizing both multi-cluster and local-cluster schedulers. Once receiving a target LLM along with its recommended configurations, the multi-cluster job scheduler will start communication with designated local clusters. Concurrently, the local-cluster job scheduler launches the LLM service on available resources, corresponding to the predefined LLM serving framework. Notably, during the service deployment, ENOVA will support various LLM technologies, including vllm~\cite{kwon2023efficient} and TensorRT-LLM~\cite{tensorrt-llm}. Following deployment, the LLM service becomes accessible through a multi-cluster ingress system. This setup ensures that incoming user requests are efficiently directed to the appropriate local cluster, where they are handled by a specific replica of the LLM service.

The monitoring system is designed for monitoring and auto-scaling, which contains real-time data collection, storage, and consumption. First, the involved data are mainly from three layers, namely hardware, LLM inference, and load balancer. For data collection in hardware and load balancer layers, ENOVA directly adopts previous mature toolkit like dcgm exporter~\cite{dcgm-exporter} and Prometheus exporters~\cite{prometheus}. For data collection in LLM inference layer, ENOVA provides a fine-grained collection tool based on opentelemetry collector~\cite{opentelemetry} in order to deconstruct the execution processes inside LLMs. Then, all collected data will be stored in time series database and consumed via stream computing framework. The performance detector is executed in streaming computing framework. Once the service quality degradation or anomalous resource utilization is detected, ENOVA will redetermine the optimal configurations and reschedule the LLM service via multi-level job scheduler.

\section{Evaluation}\label{sec:6}

In this section, we evaluate whether ENOVA effectively address the practical problems mentioned in section~\ref{sec:3.1} and accomplish the goal that developers no longer need to spend time providing stable and scalable LLM service. Experiments are conducted to investigate the following questions.

\begin{itemize}
    \item \textbf{RQ1:} Whether the service configuration module can achieve \textbf{multi-agent} and \textbf{multi-GPU deployment goals} by reducing cost and enhancing interactive performance of LLM service?
    \item \textbf{RQ2:} Whether the performance detection module can achieve \textbf{autoscaling goal} by accurately detecting anomalous resource utilization or service quality and redetermining the configurations of autoscaling?
\end{itemize}

\begin{figure*}
    \centering
    \includegraphics[width=\linewidth]{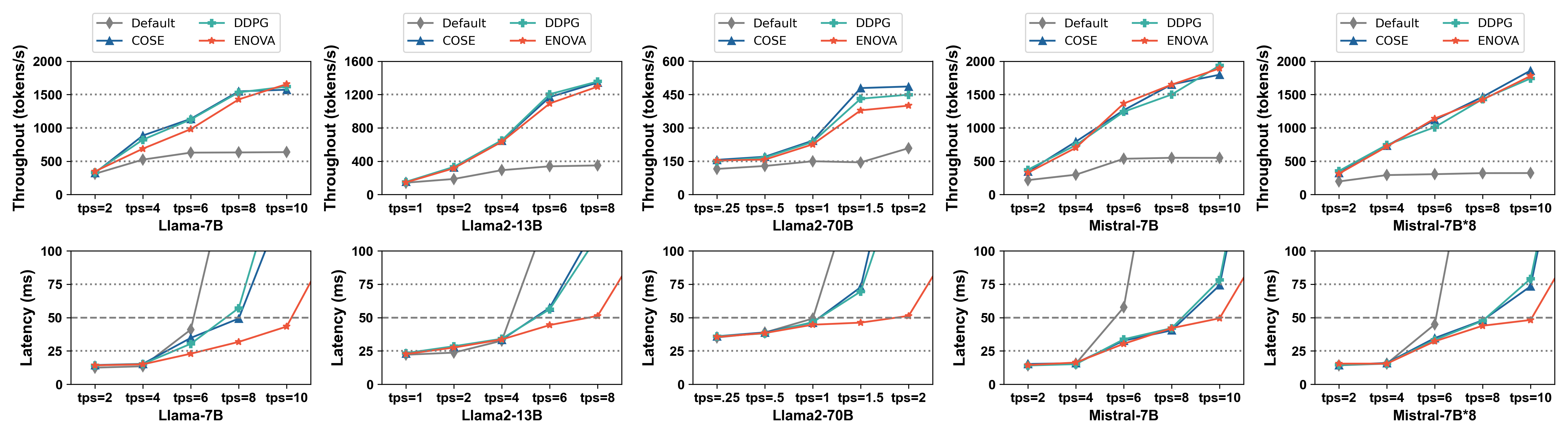}
    \caption{The throughput and latency performance comparison between ENOVA and baselines on five LLMs.}
    \label{fig:op_config}
\end{figure*}

\subsection{RQ1: Whether Service Configuration Module Enhances Performance Efficiently}\label{sec:6.2}

\subsubsection{Experiment setup}

To evaluate the service configuration module in ENOVA, we have designed experiments to assess the serving performance using ENOVA compared with established baselines. First, we selected the experimental datasets and devices to reflect the diversity of task targets and deployment in multi-GPU clusters. For this purpose, we utilize two standard datasets: grade school math 8K (gsm8k)~\cite{cobbe2021training}, representing mathematical tasks, and mostly basic python programming (mbpp)~\cite{austin2021program}, denoting programming domains. We meticulously designed the data prompts to evaluate the performance under various prompting paradigms, including zero-shot, few-shot, and chain-of-thought approaches. An example of zero-shot prompt is as follows: 
\begin{quote}
"You are a software development expert skilled in Python programming. Your task is to develop a Python function that meets the following specifications. Please ensure your code is concise, well-documented, and accurately meets all the specified requirements.\\ \\
\textbf{Task Description:} \\ 
Write a function to find the minimum cost path to reach (m, n) from (0, 0) for the given cost matrix cost[][] and a position (m, n) in cost[][]."
\end{quote}
The hardware setup includes small-scale multi-GPU clusters built with NVIDIA A100 80 GB and NVIDIA RTX 4090 24 GB, each equipped with 8 GPUs. Furthermore, to simulate realistic LLM services, we deploy LLMs with multiple architectures such as Llama2~\cite{touvron2023llama} with $7$B, $13$B and $70$B parameters and Mistral\cite{jiang2023mistral} with $7$B and $8\times 7$B parameters. The deployment of LLM services on these clusters is tested under certain load conditions, generated using a Poisson distribution for request arrival times, as outlined in~\cite{kwon2023efficient}.

Second, the baseline methods are selected to evaluate configuration recommendation module in ENOVA. Here we choose three baselines which are: 
\begin{itemize}
    \item \textbf{Default} configurations are adopted as a blank baseline, which is set to verify the necessity of configuration recommendation. 
    \item \textbf{COSE}~\cite{akhtar2020cose} is a framework that uses Gaussian Process Bayesian Optimization (GPBO) to find the optimal configurations of LLM service. 
    \item \textbf{DDPG}~\cite{lillicrap2015continuous} is a reinforcement learning technique which combines both Q-learning and policy gradients to find optimal configurations.
\end{itemize}
When adopting baseline approaches to recommend configurations, the objective function is to maximize the throughput of LLM service. In addition, due to the lack of deployment and scheduling ability of baselines, recommended configurations of baselines will be executed based on deployment execution module in ENOVA, while both the replicas of LLM service in A100 80 GB and 4090 24 GB are set to $1$ to avoid interference from hardware factors.

Third, the evaluation metrics need to characterize the serving performance and service accuracy. For serving performance,  we adopt two metrics, namely \textbf{throughput}~\cite{kwon2023efficient} which is the average number of output tokens per GPU per second and \textbf{latency}~\cite{yu2022orca} which is the average execution time of user requests divided by its output length. In addition, we adopt $15$ minute traces to evaluate serving performance in high-throughput system. For service accuracy, we adopt \textbf{accuracy}~\cite{cobbe2021training} for gsm8k dataset which is the ratio of corrected answered requests to all requests and \textbf{pass@1}~\cite{austin2021program} for mbpp dataset which is the ratio of passed answers to all requests.

\subsubsection{Overall serving performance}

We evaluate the serving performance of ENOVA and baselines on five LLMs and two datasets in different application domains. Table~\ref{tab:config_rec} shows the part of recommended configurations of ENOVA and baselines. Since baselines can not cluster user requests, we take gsm8k and mbpp as two clusters, and directly recommend the max tokens of these two datasets. We can observe that baselines have higher \textit{max\_num\_seq} and \textit{max\_tokens} for higher throughput of LLM service. Then we compare the serving performance with recommended configurations based on ENOVA and baselines. Fig.~\ref{fig:op_config} presents the throughput and latency on five LLMs with increasing requests per second (denoted as tps). First, we can observe that as the tps of user request increases, the throughput tends to the upper limit, while the latency slightly increases at first and then suddenly explodes. The is caused by that fact that when the number of user requests being processed exceeds maximal \textit{max\_num\_seq}, new requests will be pended until one of previous requests has completed. With more and more pending requests, the latency of LLM service explodes.

\begin{table}
    \centering
    \caption{The part of recommended configuration comparison between ENOVA and baselines.}
    \label{tab:config_rec}
    \renewcommand\arraystretch{1.2}
    \begin{tabular} { cccccc }
    \hline
    & LLM & {\textit{max\_num\_seq}} & {\textit{max\_tokens}} & {\textit{weights}} \\
    \hline
    \multirow{2}{*}[-0.5em]{Default}
        & L-7B & \makecell[c]{A100: 8\\4090: 8} 
                    & \makecell[c]{gsm8k: 256\\mbpp: 256} 
                    & \makecell[c]{A100: 1\\4090: 1} \\
        & L-70B & \makecell[c]{A100: 8\\4090: 8} 
                    & \makecell[c]{gsm8k: 256\\mbpp: 256} 
                    & \makecell[c]{A100: 1\\4090: 1} \\
    \hline
    \multirow{2}{*}[-0.5em]{COSE}
        & L-7B & \makecell[c]{A100: 502 \\4090: 462} 
                    & \makecell[c]{gsm8k: 970 \\mbpp: 1681} 
                    & \makecell[c]{A100: 1\\4090: 0.92} \\
        & L-70B & \makecell[c]{A100: 42 \\4090: 24 } 
                    & \makecell[c]{gsm8k: 784 \\mbpp: 1879 } 
                    & \makecell[c]{A100: 1\\4090: 0.57} \\
    \hline
    \multirow{2}{*}[-0.5em]{DDPG}
        & L-7B & \makecell[c]{A100: 452 \\4090: 386 } 
                    & \makecell[c]{gsm8k: 629 \\mbpp: 1648} 
                    & \makecell[c]{A100: 1\\4090: 0.85} \\
        & L-70B & \makecell[c]{A100: 34 \\4090: 20 } 
                    & \makecell[c]{gsm8k: 728 \\mbpp: 1938 } 
                    & \makecell[c]{A100: 1\\4090: 0.59} \\
    \hline
    \multirow{2}{*}[-0.5em]{{ENOVA}}
        & L-7B & \makecell[c]{A100: 144 \\4090: 128} 
                    & \makecell[c]{gsm8k: 414 \\mbpp: 956} 
                    & \makecell[c]{A100: 1\\4090: 0.89} \\
        & L-70B & \makecell[c]{A100: 24\\4090: 16} 
                    & \makecell[c]{gsm8k: 414 \\mbpp: 956} 
                    & \makecell[c]{A100: 1\\4090: 0.67} \\
    \hline
    \end {tabular}
\end{table}

\begin{figure}
    \centering
    \includegraphics[width=\linewidth]{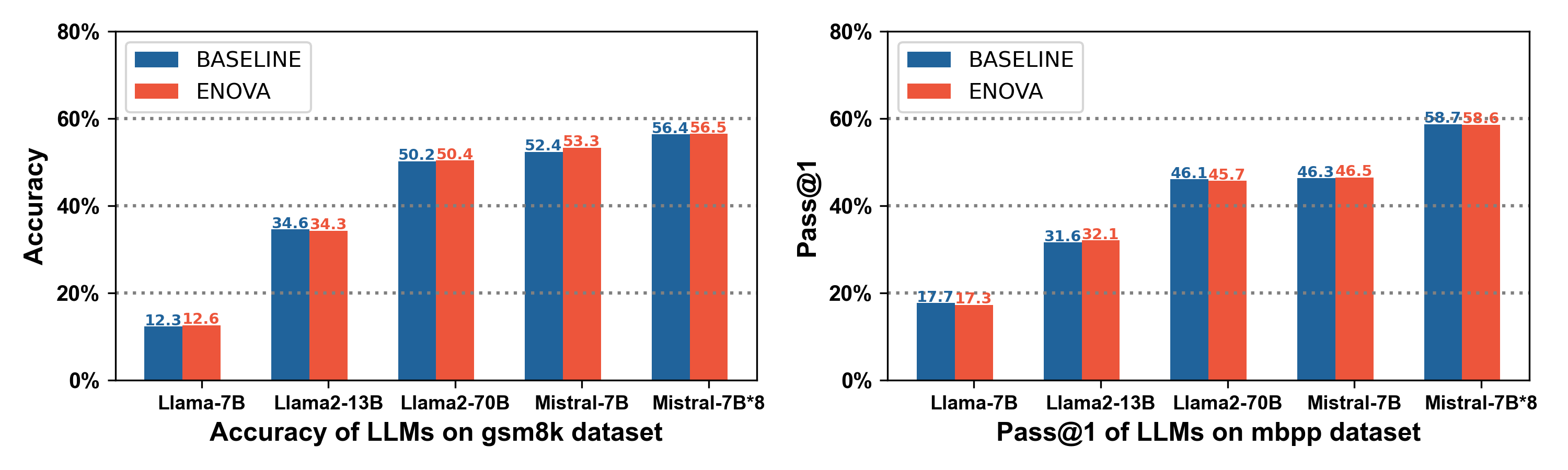}
    \caption{The accuracy and pass@1 of five LLMs on gsm8k and mbpp dataset respectively.}
    \label{fig:acc}
\end{figure}

Second, increasing \textit{max\_num\_seq} will not result in a higher throughput, but a higher latency. The upper five subplots in Fig.~\ref{fig:op_config} have presented that the throughput of ENOVA, COSE and DDPG on five LLMs has no significant differences, although COSE and DDPG recommend higher values. Further analysis reveals that increasing \textit{max\_num\_seq} leads to the processing of a greater number of user requests during next token generation, which consequently results in higher latency. It appears that the final throughput approaches a limit, which is primarily constrained by the computational capabilities of the GPUs utilized. This observation underscores the critical role of ENOVA in monitoring the execution processes of LLM services. By integrating monitoring metrics that capture the diversity of tasks assigned to LLM agents and the characteristics of different GPU devices, ENOVA can effectively map the execution dynamics of LLMs across various GPU configurations. This capability enables ENOVA to determine the optimal \textit{max\_num\_seq}. 

Third, improper \textit{max\_tokens} and \textit{weight} may cause lower tps and early explosion. As depicted in the lower five subplots of Fig.~\ref{fig:op_config}, ENOVA consistently achieved higher tps compared to baseline configurations. This improvement is attributed to ENOVA's superior handling of diverse requests. Specifically, not all user requests have optimally designed prompts that LLMs can predict accurately, leading some requests to generate the maximum allowable output tokens. An increased \textit{max\_tokens} configuration is associated with higher end-to-end execution times, which in turn results in reduced tps and potential early system overload. Moreover, an improperly configured load balancer \textit{weight} can further exacerbate these issues. If the load balancer routes new requests to already heavily loaded replicas, it can lead to early explosions even when other replicas remain underutilized. The implications of these configurations are essential, as evidenced by ENOVA's ability to sustain tps $2\times$ as high as those achieved with default configurations and $1.3\times$ higher than those of other baselines.

\subsubsection{Accuracy analysis for service configuration module}

When assessing the service configuration module in ENOVA, an essential analysis was conducted to determine if the configurations recommended by ENOVA impacts the accuracy of LLMs. Here we compare the accuracy of LLM service deployed by ENOVA with those by BASELINE, where the \textit{max\_tokens} is set to the maximum supported by the LLMs. The results, depicted in Fig.~\ref{fig:acc}, indicate no significant difference in either accuracy or pass@1 performance between the ENOVA and BASELINE. This finding illustrates that the configurations recommended by ENOVA are robust enough to support accurate output of user requests across various task targets. Consequently, the service configuration module in ENOVA has the ability to improve serving performance without affecting accuracy.


\subsection{RQ2: Whether Performance Detection Module Determines autoscaling Accurately}

\begin{table}
    \centering
    \caption{The detection performance comparison between ENOVA and baselines.}
    \label{tab:op_scaling}
    \renewcommand\arraystretch{1.1}
    \begin{tabular} { cccc }
    \hline
    & \textbf{Precision} & \textbf{Recall} & \textbf{$\mathbf{F_1}$-score} \\
    \hline
    USAD & 0.767 & 0.586 & 0.664 \\
    SDF-VAE & 0.683 & 0.829 & 0.749 \\
    Uni-AD & 0.805 & 0.753 & 0.778 \\
    ENOVA & \textbf{0.899} & \textbf{0.849} & \textbf{0.873} \\
    \hline
    \end{tabular}
\end{table}

\subsubsection{Experiment setup}
To evaluate the performance detection module in ENOVA, we design experiments to compare the detection accuracy of ENOVA and those of baselines. Due to the lack of standard datasets designed for detecting anomalies in LLM services, the experimental datasets are collected based on the chatbot service deployed in industrial environments. This service provides $8$ deployed LLMs with $2$ replicas for customers to select from. We collect the performance metrics mentioned in section~\ref{sec:4.2} of these $8$ LLMs in $4$ weeks, and adopt the collected metrics in previous $2$ weeks to train VAE-based detection model in ENOVA and baselines and those in latter $2$ weeks to test them. Hence, there are totally $1440 * 14 * 8 * 2 = 322560$ points to be detected, including 251 points labeled as anomalies.

We compare ENOVA with three multivariate time series anomaly detection baselines including  
\begin{itemize}
    \item \textbf{USAD}~\cite{audibert2020usad} adopts auto-encoder architecture and  adversarial training to isolate anomalies.
    \item \textbf{SDF-VAE}~\cite{dai2021sdfvae} designs a static and dynamic
factorized VAE to learn representations and uses reconstruction probability to identify anomalies.
    \item \textbf{Uni-AD}~\cite{he2022share} introduces transformer encoder layers and base layers to detect anomalies. 
\end{itemize}

We adopt Precision, Recall and $\mathrm{F}_1$-score to evaluate the detection performance of ENOVA and baselines. When calculating the performance, we adopt a point-adjusted approach~\cite{huang2022semi}. For any segment detected as an anomaly, if there is at least one point in the segment labeled as an anomaly, this segment is detected correctly without false positives.

\subsubsection{Overall detection performance}

When evaluating detection performance within industrial environments, the semi-supervised learning-based performance detection model of ENOVA was compared with baselines. The comparative results are outlined in Table~\ref{tab:op_scaling}, which demonstrate that ENOVA consistently outperforms the baseline models. This enhanced performance can be attributed to the design of ENOVA’s detection model, which utilizes a semi-supervised learning approach to develop more robust representations. While baseline models predominantly focus on identifying normal operational patterns, they do not adequately handle the detection of anomalies. Due to the optimized objective function, ENOVA can learn robust representations to capture the normal patterns of performance metrics, while staying away from the anomalous metrics. Overall, the semi-supervised learning framework of ENOVA can detect anomalous resource utilization or service quality accurately in industrial environments.

\begin{figure}
    \centering
    \includegraphics[width=\linewidth]{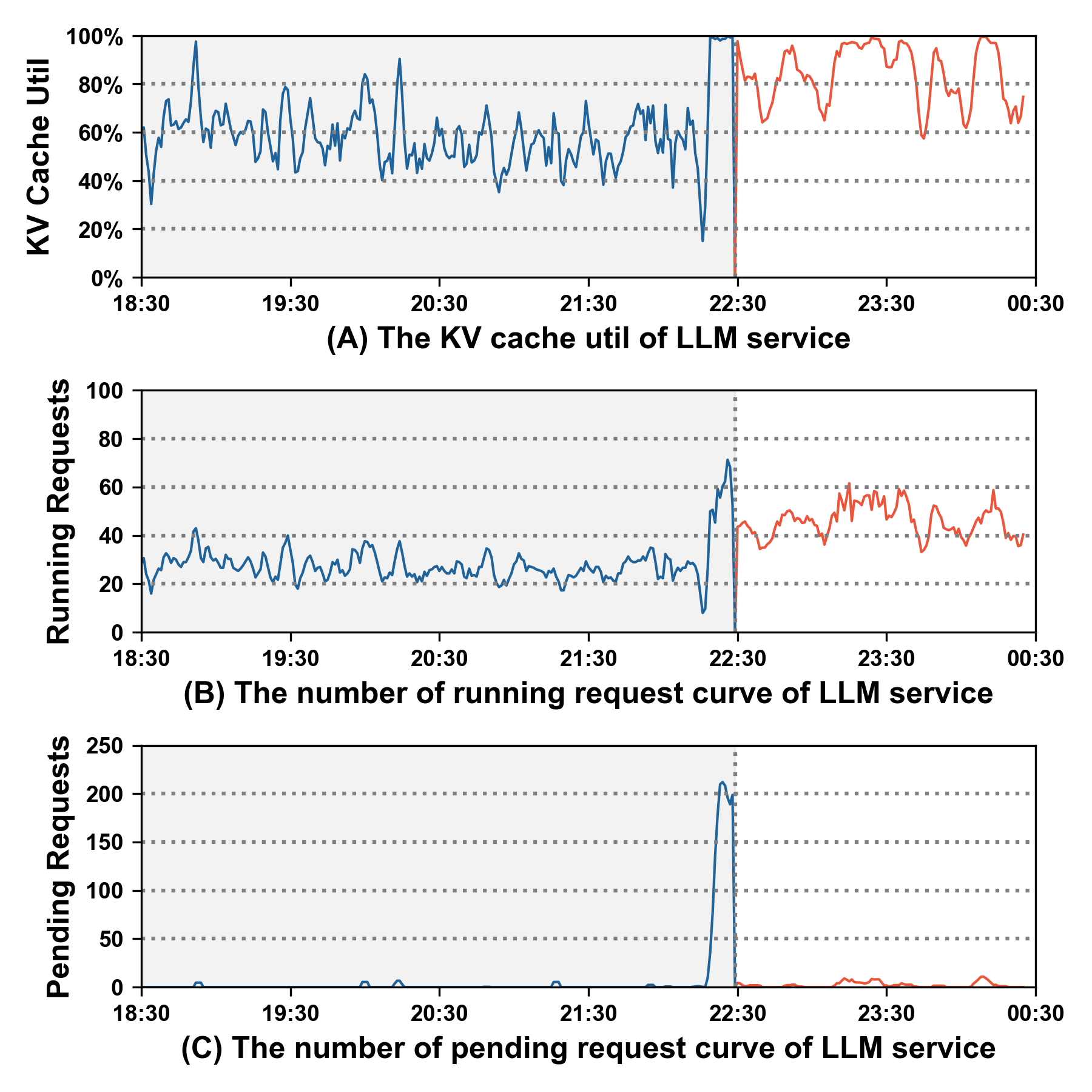}
    \caption{The KV cache utilization, number of running and pending requests of one LLM service which is one Mistral-7B deployed in RTX4090 24 GB.}
    \label{fig:autoscale}
\end{figure}

\subsubsection{Autoscaling case study}

Beyond the evaluation of detection performance, we conducted a comprehensive analysis of autoscaling efficacy by tracking the real cases in industrial environments. Figure~\ref{fig:autoscale} illustrates the performance metrics of a specific case where ENOVA effectively detect anomalous resource utilization and dynamically scaled the LLM service. In this case, Mistral-7B is deployed in one RTX4090 24 GB with default 90\% allocated GPU memory. The number of user request per second rise at 10:20, and then the KV cache utilization went to $100\%$. Without enough GPU memory for KV cache, new requests started being pended, and ENOVA detected the anomalous performance at 10:22. ENOVA can localize the lack of GPU memory for KV cache due to anomalous resource allocation, and recalculate the required GPU memory based on current requests by increasing the allocated GPU memory to $95\%$. When receiving the input to increase the allocated GPU memory, ENOVA relaunched the LLM service at 10:29. Eventually, we can observe that the LLM service can sustain about $1.6 \times$ more requests while only one configuration was adjusted without adding new replicas. Overall, this case study demonstrates the robust capability of ENOVA to ensure the stability of serverless LLM serving.

\section{Discussion}\label{sec:7}

\begin{figure}
    \centering
    \includegraphics[width=\linewidth]{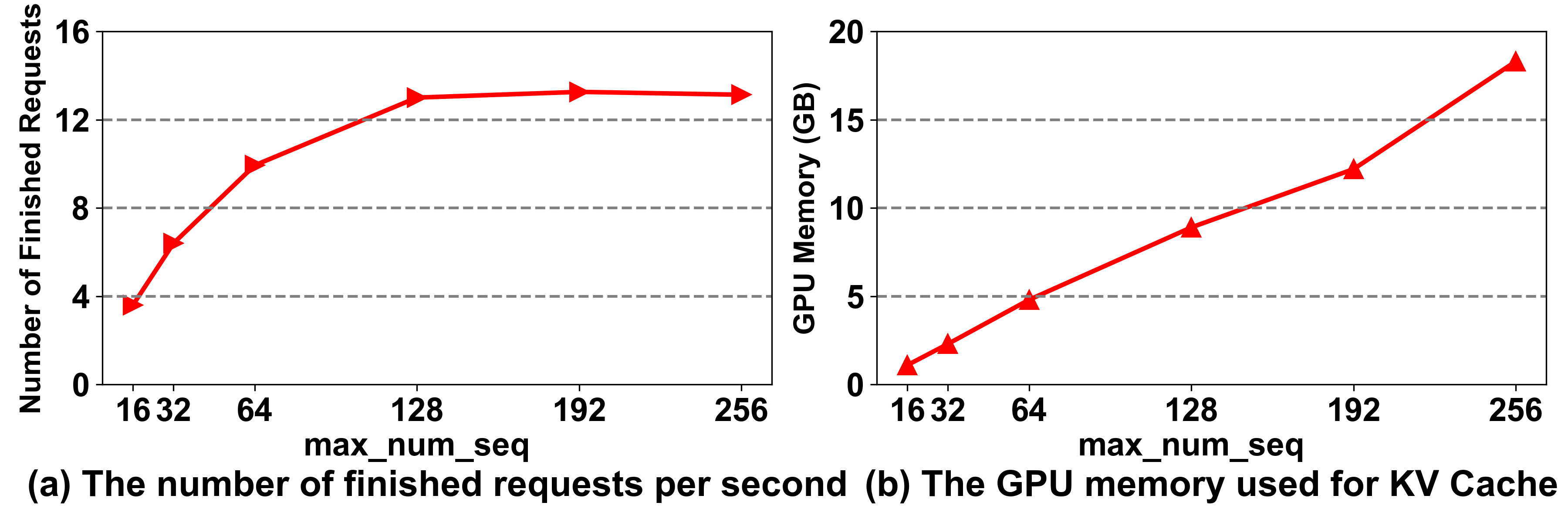}
    \caption{The maximal number of finished requests per second and GPU memory utilization for KV cache as \textit{max\_num\_seq} increases.}
    \label{fig:max_seq}
\end{figure}

\subsection{Maximal Number of Requests Processed by LLM services }\label{sec:7.1}

The maximal number of requests per second that can be processed by an LLM service is theorized to be a constant value. Suboptimal service configurations can lead to a decrease in this performance metric, while optimizing configurations will not significantly increase this metric. To investigate this hypothesis, experiments were conducted to examine the impact of varying the $max\_num\_seq$ configuration. Fig.~\ref{fig:max_seq} presents the experimental results, including both the maximal number of finished requests per second and GPU memory utilization for KV cache as \textit{max\_num\_seq} increases. Notably, the experimental data reveal that although the maximal number of finished request per second improves initially increases with higher values of $max\_num\_seq$, it eventually reaches a peak. Meanwhile, GPU memory utilization continues to rise without a corresponding increase in processing efficiency, indicating diminishing returns on resource allocation beyond a certain threshold. This observation emphasizes the critical importance of determining $max\_num\_seq$ within an optimal range to ensure maximal operational efficiency. Overall, the objective of the proposed service configuration module is to maximize the number of requests processed per second by LLM services, while minimizing GPU resource consumption.

\subsection{Processing Requests from Various LLM Agents}\label{sec:7.2}

\begin{figure}
    \centering
    \includegraphics[width=0.9\linewidth]{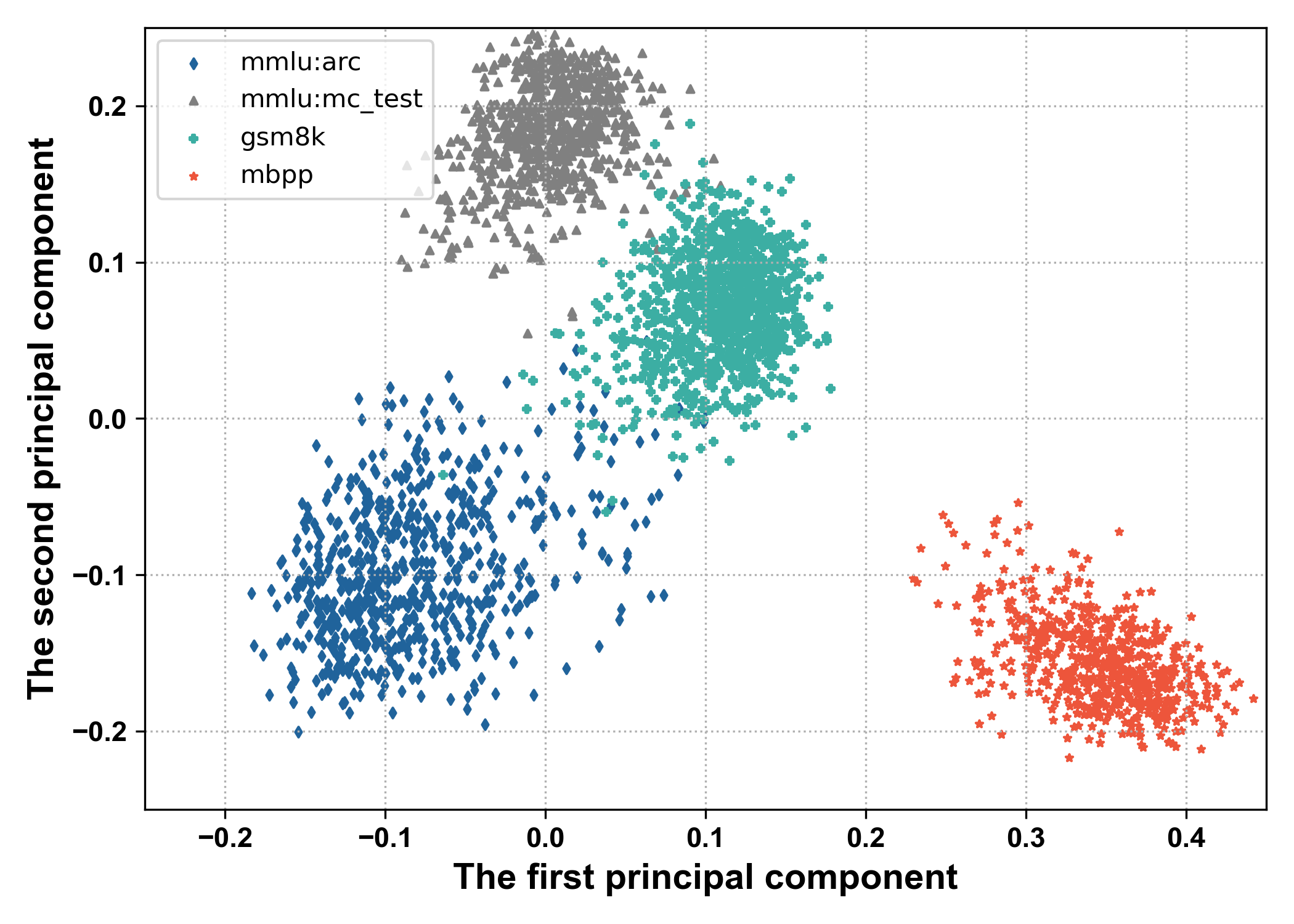}
    \caption{The principal component comparison among embedding representations of requests from different standard tasks.}
    \label{fig:domain}
\end{figure}

Requests from various LLM agents can be effectively characterized and differentiated through clustering techniques. This is critical for all LLM services, as inappropriate \textit{max\_tokens} in user requests may lead to lower tps and early explosion, as explained in section~\ref{sec:6.2}. To further analyze the difference among different LLM agents, we analyze requests from four standard datasets: gsm8k, mbpp, and ARC and MC\_TEST. For these distinct types of tasks, data prompts are crafted based on different prompting paradigms, including zero-shot, few-shot, and chain-of-thought approaches. Then, we utilize principal component analysis (PCA) to perform dimensionality reduction on the embedding representations of user requests across these tasks. Fig.~\ref{fig:domain} presents principal component comparison among embedding representations of requests from the standard tasks. The analysis reveals that requests associated with the same type of task cluster together, while those from different tasks are distinctly separated. This clustering enables us to effectively differentiate the requests based on their task types, thus enhancing the service qualities of LLM services via differentiated service configurations.

\subsection{Limitations}\label{sec:7.3}

This paper primarily proposes a deployment, monitoring, and autoscaling service aimed at improving the stability and cost-efficiency of LLM services. However, our evaluation is currently limited to specific types of LLM tasks and GPU devices. 
In the future, we aim to extend the deployment of LLMs to serve a broader range of application agents and hardware platforms, thereby enhancing both the robustness and scalability across diverse industrial environments.

\section{Related Works}\label{sec:8}

\subsection{Cloud Configuration}

Picking the right cloud configuration is essential to service quality and commercial competitiveness. Several approaches~\cite{alipourfard2017cherrypick,wang2021morphling,li2014mronline,hu2023lucid} have been proposed to manage different types of tasks run in cloud, such as MapReduce~\cite{li2014mronline}, SQL queries~\cite{alipourfard2017cherrypick}, deep learning jobs~\cite{wang2021morphling,hu2023lucid}. MROLINE~\cite{li2014mronline} designs gray-box based smart hill climbing algorithm to find the near-optimal configuration of MapReduce job. CherryPick~\cite{alipourfard2017cherrypick} leverages Bayesian optimization to build performance models and distinguishes the best configuration for data analytics jobs like SQL queries. Morphling~\cite{wang2021morphling} trains a meta-model to capture the general performance trend so that to find the optimal configuration. Lucid~\cite{hu2023lucid} adopts packing analyze model to forecast throughput, based on which the best configuration will be selected for faster deep learning job completion. However, Few of them focus on LLM service, where ENOVA is designed to settle.

\subsection{Autoscaling}

Autoscaling is the key technology to fine-grained cluster resource management for stable and cost-effective service. Many methods~\cite{qiu2020firm,rzadca2020autopilot,zhang2021sinan,bhardwaj2023cilantro} have been proposed towards different types of workloads in clusters. FIRM~\cite{qiu2020firm} detects SLO violations in microservice and takes actions to mitigate SLO violations via dynamic reprovisioning. Sinan~\cite{zhang2021sinan} also focuses on microservice autoscaling for QoS-aware resource management. Autopilot~\cite{rzadca2020autopilot} aims to reduce slack, which is the difference between the limit and the actual resource usage of workloads in kubernetes clusters. Cilantro~\cite{bhardwaj2023cilantro} introduces online learning mechanism to forms feedback loops with the jobs in cluster for performance-aware resource allocation. Inspired by these approaches, ENOVA designs performance detection module for autoscaling towards LLM service.


\section{Conclusion}\label{sec:9}

This paper proposes ENOVA, a service designed to enhance the stability and cost-efficiency of serverless LLM serving through deployment, monitoring, and autoscaling capacities. ENOVA designs a service configuration module for automatic deployment on any GPU clusters and a performance detection module for autoscaling. On top of them, ENOVA implements a deployment execution engine that manages deployment, monitoring, and autoscaling across multi-GPU clusters. We conduct experiments in various scenarios and the results show that ENOVA significantly outperform other state-of-the-art methods and is suitable for wide deployment in large online systems. In the future, we plan to expand the deployment of LLMs across a broader spectrum of application agents and hardware platforms, thereby ensuring robust performance and scalability across diverse industrial environments.

\bibliographystyle{IEEEtran}
\bibliography{reference.bib}

\end{document}